# Capping agent control over the physicochemical and antibacterial properties of ZnO nanoparticles

David Rutherford[1], Markéta Šlapal Bařinková[1], Thaiskang Jamatia[2], Pavol Šuly[2], Martin Cvek[2] and Bohuslav Rezek[1]

[1] Faculty of Electrical Engineering, Czech Technical University in Prague, Technická 2, 16227 Prague, Czech Republic
[2] Centre of Polymer systems, Tomas Bata University in Zlin, Trida T. Bati 5678, 760 01 Zlín, Czech Republic

## Abstract

Life science research has largely benefited from the use of nanoparticles (NP), yet fundamental issues such as colloidal stability and control over NP size and shape affect NP properties and functions in biomedical applications. Here we show that including capping agents directly into zinc oxide (ZnO) NP synthesis can lead to better control of these properties and their enhanced functionality. A systematic study of the influence capping agents has on the physicochemical and antibacterial properties of ZnO NP synthesized using the microwave (MW)-assisted polyol method is presented. Primary NP size (10-20 nm) was controlled by zinc precursor concentration, and NP shape was influenced by capping agent type. Capping agents enabled control over the surface charge and water interaction properties of ZnO for further investigations involving bacteria. The superior antibacterial effect was observed using positively charged, hydrophilic HMTA-capped ZnO, yet negatively charged hydrophobic OA-capped ZnO still exhibited an antibacterial effect. These observations suggest different underlying mechanisms, and we discuss these differences with particular reference to the specific surface area of ZnO, and how this is key to bacteria-nanoparticle interactions. Appropriate selection of capping agents is crucial for the synthesis of potent ZnO NPs intended for antibacterial applications, specifically for combating resistance.

**Graphical Abstract**

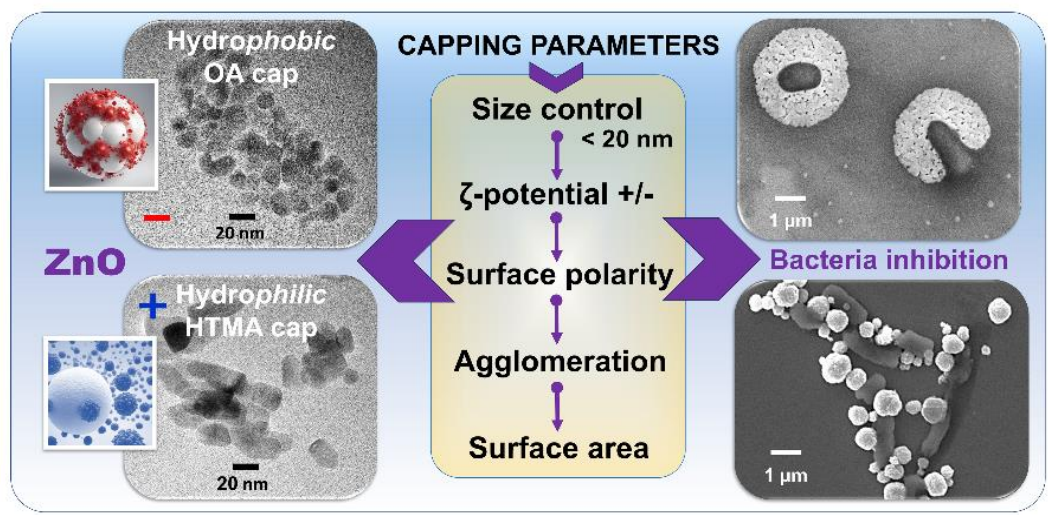

# 1. Introduction

Zinc oxide (ZnO) is a wide-band gap semi-conductor material that has been used in many applications across a range of industries owing to its unrivalled versatility, including drug delivery [1] and antibacterial applications [2], [3], [4] in the pharmaceutical industry, as well as catalytic [5], [6], energy harvesting [7], and photovoltaics [8] in the energy industry. ZnO in the form of nanoparticles (NPs) often possesses unique properties that offer enhanced activity over its bulk counterpart, further showcasing its usefulness as a multi-functional material. The relative ease with which ZnO NPs can be synthesized only increases their prevalence, from gas-phase plasma synthesis [9] to wet chemical-based methods including sol-gel [10], co-precipitation [11], hydrothermal growth [12], [13], and microwave (MW)-based systems [5], [14], [15].

Advancements in nanotechnology have enabled the widespread synthesis of NPs for use as colloidal suspensions, where the stability of synthesized NPs depends heavily on the solvent type and ionic concentration [16], [17]. Molecules that bind to the surface of NPs, so-called capping agents, can alter the properties of the NPs and improve their stability and solubility [11], [18]. Other properties, such as reactivity and biocompatibility, can also be influenced by capping agents. For example, a hydrophilic capping agent like hexamethylenetetramine (HMTA) can improve nanoparticle affinity in aqueous environments, whereas a hydrophobic capping agent like oleic acid (OA) would enhance stability in non-polar solvents. Furthermore, capping agents can influence the surface charge, surface chemistry, and interactions with other molecules or cells. The addition of capping agents allows for the precise control and optimization of nanoparticle properties, making them more suitable for a wide range of applications. However, it is important to optimize the concentration of the capping agent and synthesis conditions in order to achieve the required properties of the NPs [19].

The application of capping agents plays a crucial role in the synthesis of the ZnO NPs, as they allow for controlling the size, morphology, and stability in various media. Recently, different materials (organic ligands, polymers, dendrimers, amino acids, etc.) have been explored as capping agents to tailor the properties of ZnO NPs [20]. **Table 1** briefly summarizes the range of different techniques that have been used to synthesize ZnO NPs in the presence of various capping agents. For instance, Tang et al. synthesized ZnO nanorods via the hydrothermal route using polyvinyl alcohol (PVA), polyethylene glycol (PEG), sodium dodecyl sulfate (SDS), and cetyltrimethyl ammonium bromide (CTAB) as capping agents [12]. They found that PVA-stabilization provided the most uniform shape of the ZnO nanorods, whose aspect ratio was further tuned by the reaction temperature, while PEG, SDS, and CTAB provided irregular-shaped nanorods with varied distributions. Akhil et al. employed the co-precipitation method to synthesize ZnO NPs capped with ethylene glycol (EG), gelatin, PVA, and polyvinylpyrrolidine (PVP) and examined their photocatalytic, antibacterial, and antibiofilm activities [11]. They reported that the presence of capping agents decreased the catalytic efficiency of ZnO due to decreased absorption of light, especially in the case of ZnO-PVA. Similarly, all the capped ZnO variants, except for PVP capping, exhibited lower antibacterial and antibiofilm activity compared to the neat ZnO. Such effects were attributed to the differences in adsorption properties and the hindering effects of the capping layer preventing electron-hole recombination. On the contrary, capping with 2-mercaptoethanol decreased the size of ZnO aggregates, which yielded enhanced antimicrobial activity as a result of the higher surface-to-volume ratio [21]. In specific cases, the hindering effects of the capping agents are desired. Cao and Zhang used tetraethyl orthosilicate (TEOS) and dimethyldiethoxysilane (DMDES) to achieve photocatalytic deactivation and improved compatibility with organic matrices for ZnO formulations applied in the cosmetics industry [22]. In the field of electronics, the ZnO NPs and their variants doped with Fe were stabilized with oleic acid (OA), yielding a stable and active material for the fabrication of a polymer light-emitting diode (PLED) [14]. In this regard, surfactants such as hexamine, tetra ethyl ammonium bromide (TEAB), CTAB, and tetraoctyl ammonium bromide (TOAB) were investigated as well, showing a reduction in the visible emission spectra [23]. Besides the organic substances, bio-sourced species, such as citrus extracts [10], potato starch [24], and carboxymethyl cellulose (CMC) [25], among other substances, have also been explored as effective stabilizers to reduce the aggregation of the ZnO NPs.

In the current study, we compare the influence of two different capping agents with opposite water interaction properties. Hydrophilic HMTA and hydrophobic OA were added separately to two different concentrations of zinc acetate dihydrate (ZAD) prior to irradiation by microwaves. The resulting NPs were thoroughly characterized using microscopic as well as spectroscopic methods, and their antibacterial effect was assessed using the standard micro-broth dilution technique to ascertain the minimum inhibitory concentration (MIC). Considering the current situation with antimicrobial resistance (AMR) and the reducing efficacy of current antibiotics against bacterial infection, the antibacterial properties of ZnO have been the subject of deep interest [26]. The mechanism through which ZnO is thought to exert its antibacterial effect is not one single process but an interplay between several different processes that may occur simultaneously. We show that, under certain conditions, ZnO synthesized in the presence of a capping agent can enhance the antibacterial effect.

*Table 1. Overview of the synthetic methods, capping agents, and relevant properties of the ZnO NPs with diverse target applications reported in the literature.*

| Synthesis technique | Capping agent | ZnO shape | ZnO dimensions | Targeted application | Reference |
|---|---|---|---|---|---|
| Sol-gel | HMTA, CTAB | Non-uniform, spherical | Concentration dependent, 0.15–0.35 μm | Biomaterials | [10] |
| Co-precipitation | EG, gelatin, PVA, PVP | Hexagonal | 52–91 nm based on capping | Photocatalysis, antibacterial | [11] |
| Hydrothermal | PVA, PEG, SDS, CTAB | Nanorod | 100 nm diameter, 3–5 μm length | Catalysis | [12] |
| MW-synthesis | OA | Oblate | 10 nm (TEM) | PLED device | [14] |
| Precipitation, base hydrolysis | 2-mercaptoethanol | Nearly spherical | 12 nm / 47 nm, with / without capping | Antibacterial | [21] |
| Commercial ZnO | TEOS, DEDMS | Irregular spheres | 10–50 nm (TEM) | Cosmetics | [22] |
| Solvothermal | Hexamine, TEAB, CTAB, TOAB | N.s. | N.s. | Luminescent devices | [23] |
| Reduction and calcinations | Starch (potato) | Relatively spherical | 100–300 nm, 20 nm (TEM) | Green nanotechnology | [24] |
| Precipitation | CMC | Irregular agglomerates | 40–50 nm | Food industry | [25] |

## 2. Methods
### 2.1. Synthesis of ZnO NPs

To fabricate ZnO NPs, a facile MW-assisted polyol synthesis reported by Hammarberg was performed [27] with some modifications. The zinc acetate dihydrate (ZAD) (p.a.) serving as a Zn precursor was placed into a Teflon-lined reactor and vigorously mixed (5 min, 1000 rpm, RT) with 50 mL of diethylene glycol (DEG) (≥99.0 %, Reagent Grade) that represented a suitable medium. A capping agent – oleic acid (OA) (≥99.0 %, 0.660 g, 2.3 mmol) or hexamethylenetetramine (HMTA) (≥99.0 %, 0.328 g, 2.3 mmol) – was subsequently added, and the mixture was stirred for an additional 10 min. The assembly was placed into a MW reactor (Magnum II, Ertec, Poland) and tightly closed. The power of the MW reactor was set to 100 % with a heat time of 15 min. The reactor was programmed so that the temperature and pressure increased and plateaued at around 245–250 °C and 45–50 bar, respectively. Afterwards, the reactor vessel was allowed to cool down to 45 °C, before being opened. The product was separated by centrifugation (EBA 21, Hettich, Germany) (5 min, 6000 rpm), washed in several cycles (3×50 mL) with methanol (≥99.6 %), and dried at 65°C (150 mbar) overnight. The reference ZnO NPs were prepared without using any capping agent. The chemicals were sourced from Sigma (USA) and Penta (Czech Republic).

To design the ZnO NPs of a desired size, the molar concentration of ZAD was changed in a range from 4 to 16 mmol, while the concentration of a capping agent and the volume of the medium were kept constant. Based on the data, a regression model (**Figure S1**) was derived to correlate the concentration of ZAD with the size of ZnO crystallites for each capping agent.

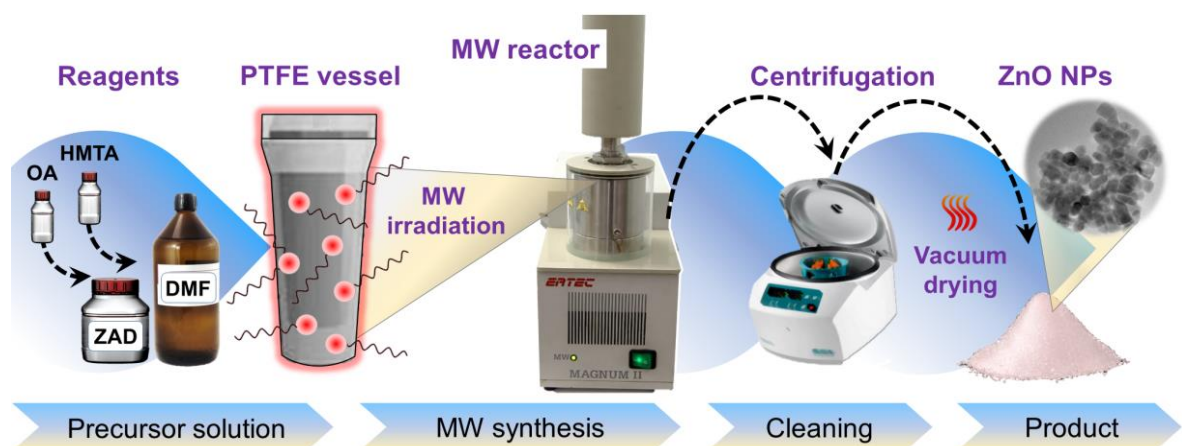

*Figure 1.* Schematic diagram showing the MW-synthesis protocol for ZnO NPs in the presence of OA and HMTA serving as capping agents.

### 2.2. Characterisation of ZnO powders

Electron microscopy was used to determine the dimensions and morphology of the ZnO primary NPs (transmission electron microscopy, TEM, JEM-2100Plus, JEOL, Japan) on a copper grid at an acceleration voltage of 200 kV, and at least 100 individual particles were manually measured using ImageJ software (National Institutes of Health, USA) to determine the average size of the primary NPs. Energy-dispersive X-ray (EDX) spectroscopy that was integrated into a scanning electron microscope (NanoSEM 450, FEI, USA) operating at an acceleration voltage of 15 kV was used to analyse the elemental composition. The crystallographic profile of the synthesized ZnO NPs was interrogated by X-ray diffraction (XRD, Miniflex 600, Rigaku, Japan) spectroscopy with a Co-Kα radiation source (λ = 1.789 Å). Finally, the specific surface area (SSA) of the synthesized ZnO NPs was determined using the Brunauer-Emmett-Teller (BET) technique from analysis of Nitrogen adsorption/desorption isotherms recorded on a volumetric gas adsorption analyzer (Belsorp Mini II, Japan) at 77 K.

*2.3.* Characterisation of ZnO dispersions

The polarity of the synthesized ZnO NPs as a colloid was determined by measuring the zeta potential (ζ-potential) using a zetasizer (Zetasizer Nano, Malvern Panalytical, UK), and particle agglomeration profiles were recorded by dynamic light scattering (DLS) analysis using the same machine in HPLC H2O (Carl Roth, Germany). A commercial assay (Amplite, AAT Bioquest, California, U.S.A.) was used to measure the concentration of $Zn^{2+}$ from the supernatants of colloidal suspensions after centrifugation to remove NPs in HPLC H2O by UV-vis spectroscopy (Epoch 2 Microplate Spectrophotometer, BioTek, Agilent, USA). The same machine was also used to detect the absorption edge of colloidal ZnO NPs synthesized in the presence of HMTA and OA capping agents.

2.4. Bacteria treatment and antibacterial analysis

2.4.1. Bacteria preparation

*Escherichia coli* (*E. coli* ATCC 25922) and *Staphylococcus aureus* (*S. aureus* ATCC 25923) were purchased from the Czech collection of Microorganisms (CCM, Czech Republic) and resuscitated at 37 °C for 24 h using orbital rotation (150 rpm, PSU-10i, BioSan, Latvia) in 100 mL Mueller Hinton Broth (MHB, Sigma Aldrich). Stock vials were prepared, containing 2 mL of the resulting culture and 1 mL of sterile glycerine for storage at -20 °C. A fresh vial was used for each experiment and allowed to thaw before a 1:10 dilution series was performed using sterile 0.9% NaCl (Penta, Czech Republic). A volume of 500 μL from each dilution was added to a Petri dish containing Mueller Hinton agar (MHA, Sigma Aldrich) and incubated at 37 °C for 24 h. A single colony was removed, added to 5 mL of MHB, and placed in an orbital shaker at 37 °C for 24 h to create a clonal population of bacteria needed for the experiments.

2.4.2. Antibacterial effect

The antibacterial effect of test samples was assessed using the industry-standard known as the minimum inhibitory concentration (MIC, EUCAST [28]) modified for use with NPs such as ZnO. Clonal populations of bacteria were adjusted to McFarlands Density 1.0 using a densitometer (Den-1, BioSan, Latvia), which is approximately equivalent to $3 \times 10^8$ colony forming units per millilitre (cfu/mL). Clonal populations were diluted a further 1000 times using MHB (i.e., 3 × 1:10 dilutions) before adding 50 μL to individual wells of a 96-well microplate that contained the same volume of different concentrations of test samples. The microwell plate was added to a spectrophotometer (Epoch 2 Microplate Spectrophotometer, BioTek, Agilent, USA) set to 37 °C and the optical density (OD) of the bacteria-nanoparticle suspension was then monitored every 15 minutes for 18 hours. The difference in the OD value at the start of the experiment and at the end of the experiment was used to negate the influence of the presence of test samples in the suspension during OD measurement ($\Delta OD_{600}@t_{18}-t_0$).

2.4.3. Observation of bacteria-ZnO interaction in suspension

An overnight clonal population of *E. coli* was washed thrice by centrifugation to remove any residual broth (13,000 rpm for 5 min) before being adjusted to MF1.0 in water. A 1:10 dilution was made before the addition of either ZnO-HMTA-16 or ZnO-OA-8 (final concentration = 100 ug/mL). The bacteria-ZnO suspensions were then incubated (1 h, 37°C, 150 rpm), before a 10 μL droplet was dropcasted onto a pre-cleaned silicon wafer for the SEM imaging at 5 kV acceleration voltage, a spot size of 10 pA, and a working distance of approx. 7.5 mm (Evo 10, Zeiss, Germany).

# 3. Results

## 3.1. Structural and dimensional analysis

The ZnO NPs were prepared by MW-assisted synthesis, which involved optimizations with various concentrations of the ZAD precursor and the application of OA and HMTA as different capping agents. It was found that the ZAD concentration efficiently controlled the size of the crystallites, as its increasing concentration resulted in the gradually increasing size of ZnO crystallites, regardless of the capping agent (**Figure S1**). The following ZnO batches were selected for further investigation: ZnO-OA-4, ZnO-OA-8, ZnO-HMTA-8, and ZnO-HMTA-16. (n.b., the number denotes the concentration of ZAD used in mmol). As seen in **Figure 2A**, the ZnO NPs were generally uniform and oblate, but the application of HMTA yielded a more elongated shape. The dimensions of primary particles for ZnO-OA-4 and ZnO-HMTA-8 were calculated based on image analysis to be 8.8 and 9.1 nm, while those for ZnO-OA-8 and ZnO-HMTA-16 were around 14.3 and 28.8 × 15.7 nm (considered as a rod). The representative image analysis is shown in **Figure S2**. Overall, it can be stated that the ZAD concentration affected the size and the type of surfactant affected the shape of the ZnO.

The XRD spectra for the investigated ZnO batches (**Figure 2B**) showed the diffraction peaks centred at 2θ angles of 37.1, 40.1, 42.3, 55.8, 66.9, 74.5, 78.9, 80.9, and 82.4°, which are related to (100), (002), (101), (102), (110), (103), (200), (112), and (201) crystallographic planes typical for the hexagonal wurtzite structure [5]. The average crystallite size, $D$, was calculated according to the Scherrer equation:

$$D = \frac{K\lambda}{\beta \cos\theta}$$

where $K$ is the Scherrer's constant (a typical value is 0.9), $\lambda$ is the wavelength of the X-ray source, $\lambda(CoK\alpha_{1,2})$ = 1.7903 Å, $\beta$ is the full width at a half-maximum (FWHM) intensity of a peak observed at a mean scattering 2-theta angle (expressed in radians), and $\vartheta$ is the Bragg angle [29]. The $D$ values for the ZnO-OA-4 and ZnO-OA-8 were 10 and 15 nm, and for the ZnO-HMTA-8 and ZnO-HMTA-16, they were 11 and 18 nm, respectively, which are values close to the size of the individual NPs, observed by TEM (**Figure 2A**).

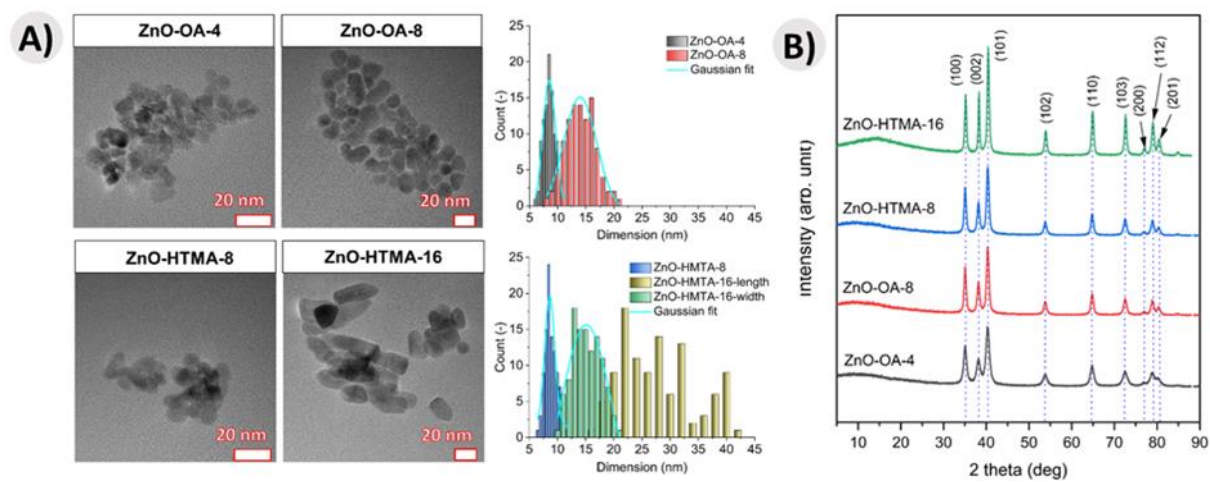

*Figure 2.* TEM images with size distributions of the primary particles (A) and XRD patterns (B) for the ZnO NPs prepared using OA and HTMA as capping agents and different amounts of the ZAD precursor.

## 3.2. Morphological analysis and elemental composition

Clear differences in the degree of agglomeration were observed from the SEM micrographs. As seen, the particle agglomeration was more pronounced for ZnO-OA that formed irregular-shaped agglomerates when compared to ZnO-HMTA that yielded smaller spherical ZnO agglomerates (**Figure 3A** vs. **3B**). It should be mentioned that solvent-based synthetic methods typically provide ZnO with inactive surfaces due to fouling of the attached organic residues, and their removal is time-consuming and requires harsh conditions [30]. For this purpose, the EDX method was applied to study the surface chemistry of the synthesized ZnO. As seen in **Figure 3C**, the EDX spectrum of ZnO-OA revealed the presence of carbon, which can be associated with the OA molecules adhered to the ZnO surface. On the contrary, the spectrum for ZnO-HMTA (**Figure 3D**) was devoid of carbon and nitrogen, which are elements constituting the HMTA molecule. This result indicates that the MW-assisted synthesis in the presence of HMTA provided ZnO with greater purity, and no additional purification treatment was needed.

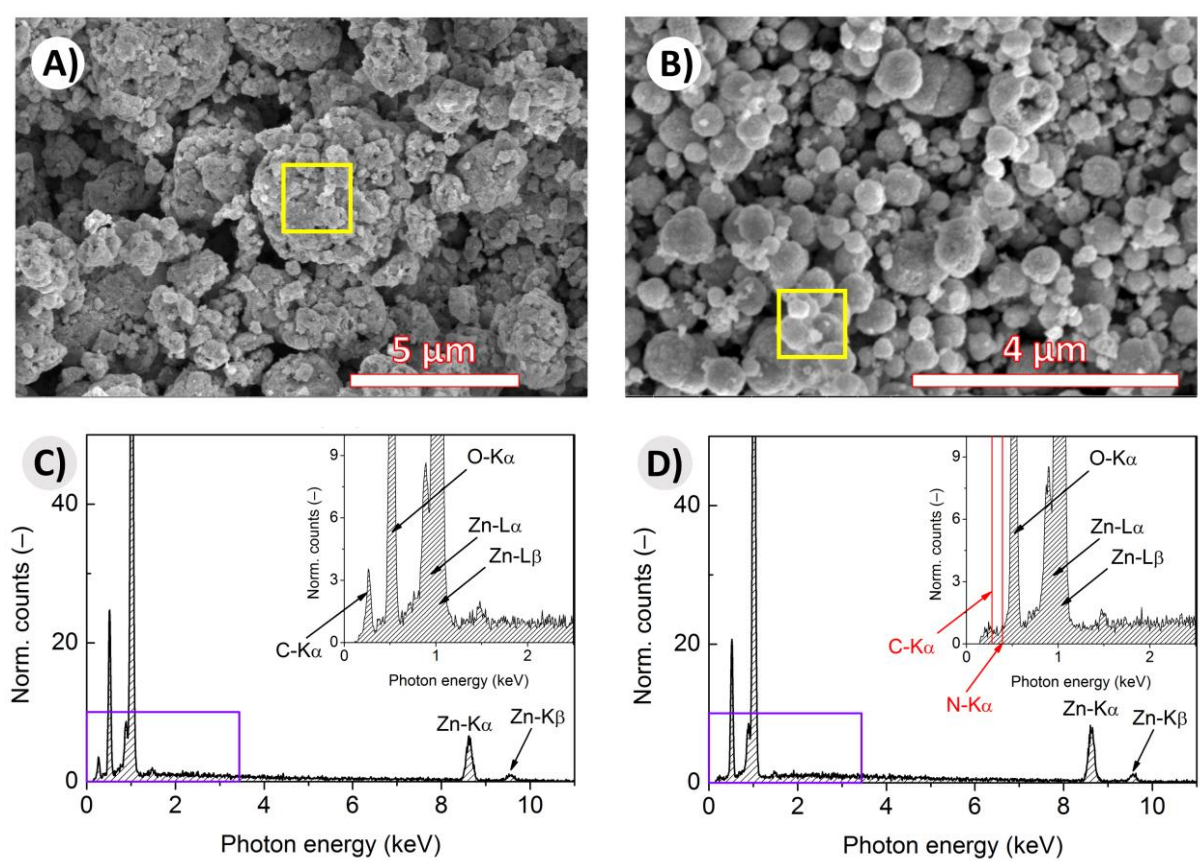

*Figure 3. SEM micrographs of ZnO-OA-8 (A) and ZnO-HMTA-8 (B), and the corresponding EDX spectra (C, D). The red lines in the latter spectrum show the absence of HMTA-related elements.*

## 3.3. Specific surface area of synthesized ZnO

As known [31], the SSA can be related to the antibacterial activity of ZnO. For this reason, the SSA was determined from the physisorption data using the BET method (**Figure S3**). The ZnO-OA-4 and ZnO-OA-8 exhibited SSA values of 0.72 and 3.50 $m^2/g$, respectively. These data suggest significant agglomeration, likely combined with the chemical aggregation [32] of these powders. In the case of ZnO-HMTA-8 and ZnO-HMTA-16, the SSA values were 42.3 and 31.8 $m^2/g$, respectively; thus, they exhibited an anticipated trend (a lower SSA for larger NPs) and were significantly higher compared to their OA-capped analogues. These findings corroborate well with the agglomeration phenomena observed by SEM micrographs on the same dry powder (**Figure 3A, B**). The possibility that this greater SSA measured from dry ZnO-HMTA powder might transfer to a larger contact area between the ZnO NPs and bacteria in the liquid phase will be investigated in the following sections.

### 3.4. Analysis of ζ-potential and ZnO size distribution in dispersions

The surface charge of synthesized ZnO samples was assessed by measuring the ζ-potential of their colloidal suspensions (**Figure 4A**). ZnO synthesized without a capping agent, serving as a reference, possessed a ζ-potential value of +35 mV. The addition of HMTA at both concentrations of ZAD yielded a slight reduction in the positive value (+23 mV), whereas OA caused a reversal of the charge to negative values. In the presence of OA, the lowest concentration of ZAD resulted in a ζ-potential value of -9 mV, yet the higher ZAD concentration registered two peaks centred at -6 and +7 mV, respectively. This suggests ZnO-OA-8 was composed of two populations of NPs with differing surface charges. All other test samples, including the uncapped ZnO, exhibited a monomodal peak in the ζ-potential spectra, which implies a single, uniform population of NPs with identical surface charge. In general, ζ-potential measurements below ±30 mV are considered unstable. In our case, surprisingly, the most stable material was ZnO synthesized without any capping agent, and the addition of both capping agents detrimentally influenced the stability of the ZnO colloidal dispersions.

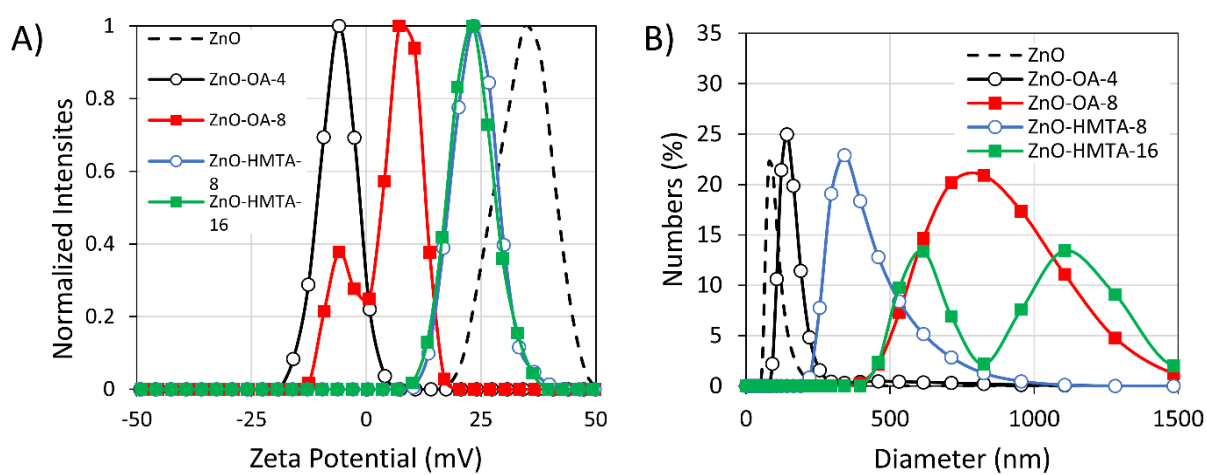

*Figure 4.* ζ-potential (A) and DLS (B) data from the tested ZnO colloidal suspensions.

**Figure 4B** details the results of the DLS analysis that was used to estimate the particle diameters of test samples as colloids (see **Figure S4** for complete particle size distributions). The smallest ZnO particle diameter of around 100 nm was measured from the uncapped ZnO, followed by ZnO-OA-4 with an average hydrodynamic diameter of 164 nm. These two test samples also recorded a second broad peak centred at approximately 750 nm. The higher concentration of ZAD in the presence of OA, i.e., ZnO-OA-8, resulted in a monomodal particle distribution with an average particle diameter of around 1000 nm. The same concentration of ZAD with HMTA (i.e., ZnO-HMTA-8) yielded particle diameters of approx. 500 nm, and doubling of its concentration resulted in the ZnO distribution with two distinct peaks at 615 and 1100 nm, respectively.

Computational studies revealed that HMTA influences the aggregation of water molecules immediately surrounding the molecule, offering insights into how capping agents exert their influence on the surrounding media as well as on the nanoparticle itself [33].

### 3.5. Optical absorption analysis (UV-Vis)

The absorbance spectra from ZnO-OA and ZnO-HMTA in the range of 300-450 nm are shown in **Figure 5** (see **Figure S5** for the full UV-vis absorbance spectra (200-900 nm). The absorbance peak relates to the absorption band edge, where photons of incident light have sufficient energy to excite electrons of the Zn-O bond, and both HMTA-capped ZnO test samples displayed a greater defined absorbance peak compared to OA-capped ZnO. Except for ZnO-HMTA-8, the peak was only detected at relatively high concentrations of test samples (i.e., 512 µg/mL). The area under the absorbance peak was greater when a higher concentration of ZAD was used during the synthesis process for both capping agents. This could be a result of a greater amount of ZnO having been synthesized when a higher concentration of precursor is added compared to lower precursor concentrations.

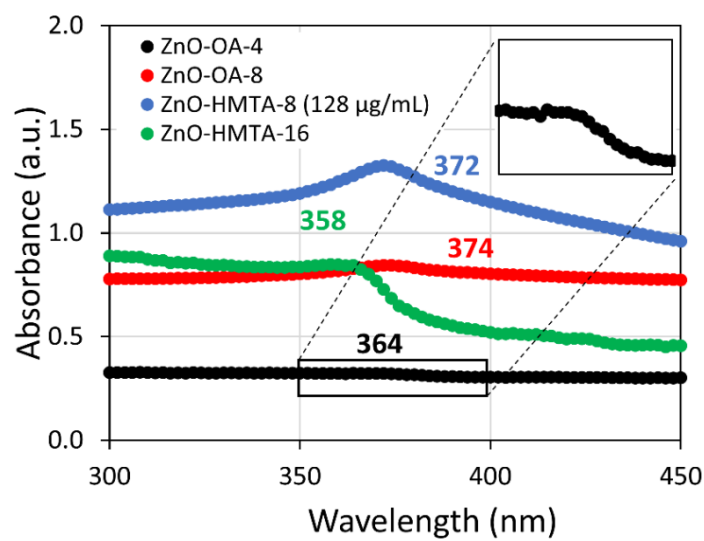

*Figure 5.* The absorbance spectra of colloidal test samples (512 µg/mL unless indicated) from 300-450 nm to observe the typical ZnO absorption peak. The inset square displays the ZnO-OA-4 peak over a narrower range to better show the absorption peak, and color-coded numbers indicate the wavelength that the peak absorbance was measured at from each test sample. The full UV-vis spectra (200-900 nm) from all concentrations of all test samples is provided in **Figure S5**.

### 3.6. Analysis of released zinc ion ($Zn^{2+}$)

The concentration of $Zn^{2+}$ measured from the supernatant of colloidal suspensions of test samples is displayed in **Figure 6** using a commercial assay (a calibration curve is shown in **Figure S6**). $Zn^{2+}$ is formed from the partial dissolution of the test samples upon interaction with water. The lowest concentration of $Zn^{2+}$ was measured from ZnO-HMTA-8 immediately prepared (0 h), which was statistically significantly lower than the other test samples for the same time interval. After 24 h storage at room temperature, the test samples were re-measured, and there was an increase in $Zn^{2+}$ detected from ZnO-HMTA-8, whereas the remaining test samples recorded similar concentrations that were measured at 0 h.

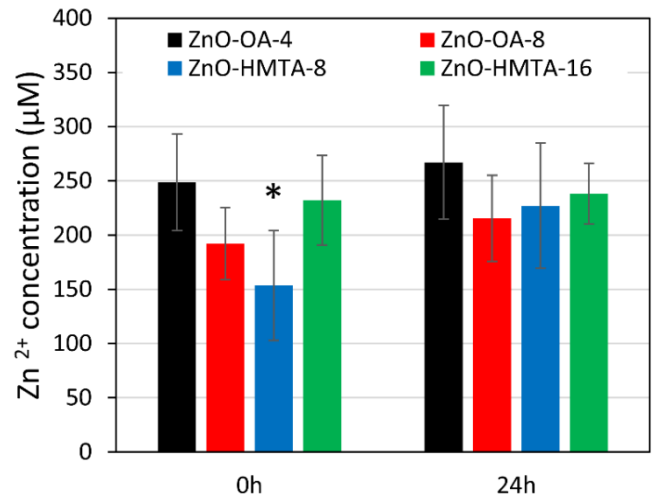

*Figure 6.* *Zinc ion concentration measured from the colloidal suspension of the test samples immediately after preparation (0 h) and after 1 day of storage at room temperature (24 h). \*=p<0.05.*

### 3.7. Minimum Inhibitory Concentration

The antibacterial effect of the test samples was assessed using the micro-broth dilution technique to discover the MIC values. The optical density (OD) of the incubated bacteria-test sample suspensions was monitored to reveal the lowest concentration that did not record an increase in OD and therefore inhibited growth. The final OD value at the end of the experiment (i.e., $OD_{600}@t_{18}$) was corrected using the initial OD value (i.e., $OD_{600}@t_0$) to account for the presence of test samples. These data were then plotted as a function of test sample concentration, and the results can be seen in **Figure 7**. Graphs showing raw OD (i.e., uncorrected) data can be seen in the supporting information (**Figure S7**). Test samples that achieve complete inhibition using the lowest concentrations are clearly the most antibacterial (i.e., when $\Delta OD_{600}@t_{18}-t_0=0$).

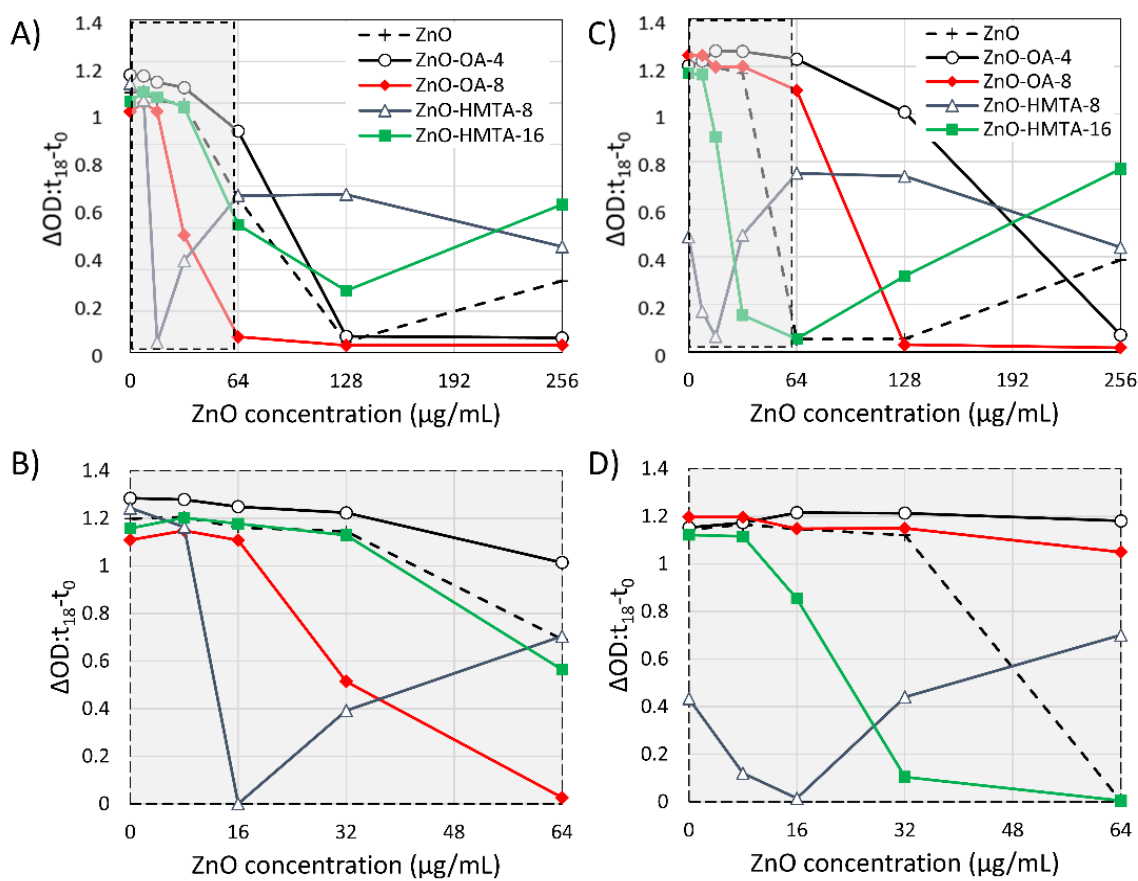

***Figure 7.*** *Differences between the OD values at 600 nm at the start and end of incubation ($\Delta OD{:}t_{18}\text{-}t_0$) as a function of test sample concentration: (A) E. coli; 0–256 µg/mL, (B) E. coli; 0–64 µg/mL, (C) S. aureus; 0–256 µg/mL, (D) S. aureus; 0–64 µg/mL.*

**Figure 7A** presents the blank-corrected OD values for *E. coli* across the entire range of test sample concentrations (0–256 µg/mL), and the shaded area magnified in **Figure 7B** showcases the region below 64 µg/mL. One can see that ZnO-HMTA-8 completely inhibited *E. coli* growth at 16 µg/mL. The OD increase observed for greater concentrations can be attributed to intensified ZnO NP sedimentation with concentration over time. ZnO-OA-8 inhibited *E. coli* growth at 64 µg/mL, with the remaining capped test samples inhibiting growth at 128 µg/mL, which was similar to the uncapped test sample. It is therefore obvious that ZnO capped with HMTA or OA significantly enhanced the antibacterial effect of ZnO at particular concentrations. As obvious from **Figure 7B**, ZnO-OA-4 exhibited a higher MIC than the uncapped ZnO, which could be explained by the hydrophobic nature of OA, which makes it difficult to create homogeneous particle dispersions. In this sense, it was observed that stock solutions of ZnO-OA exhibited sedimentation within minutes of creation (**Figure S5**). For these reasons, the concentrations derived from this stock suspension may not be accurate since the particles existed in larger agglomerates that were not suspended uniformly throughout the suspension.

The analogous interpretation was used for the corrected final OD values for *S. aureus* (**Figure 7C, 7D**). Interestingly, the test sample that inhibited *S. aureus* growth at the lowest concentration was again ZnO-HMTA-8 at 16 µg/mL, which was followed by ZnO-HMTA-16 that inhibited *S. aureus* growth at 64 µg/mL, similarly to the uncapped ZnO. OA-capped ZnO variants required concentrations above 64 µg/mL to inhibit *S. aureus* growth. As discussed above, this can in part be explained by the difficulty of creating a uniform suspension of OA-capped ZnO in the aqueous environment. It is noteworthy that ZnO-OA-8 inhibited the growth of *E. coli* at 64 µg/mL, yet *S. aureus* required a much greater concentration (256 µg/mL) to effectively inhibit its growth. This indicates that the antibacterial effect of OA-capped ZnO can be influenced by the bacteria strain and differences in the outer cell wall structures.

### 3.7.1. Interaction of ZnO nanoparticles with bacteria

The SEM micrographs of *E. coli* incubated with ZnO-OA-8 (**Figure 8A**) and ZnO-HMTA-16 (**Figure 8B**) showed remarkable differences in their interaction behaviour. Hydrophobic ZnO-OA-8 formed large islands of clustered NPs around the bacteria, whereas hydrophilic ZnO-HMTA-16 formed much smaller agglomerates, and individual particle interaction was observed. These observations suggest that the mechanisms which underlie the antibacterial effect of ZnO-OA-8 must be different from that of ZnO-HMTA-16 due to the different interaction dynamics in an aqueous environment.

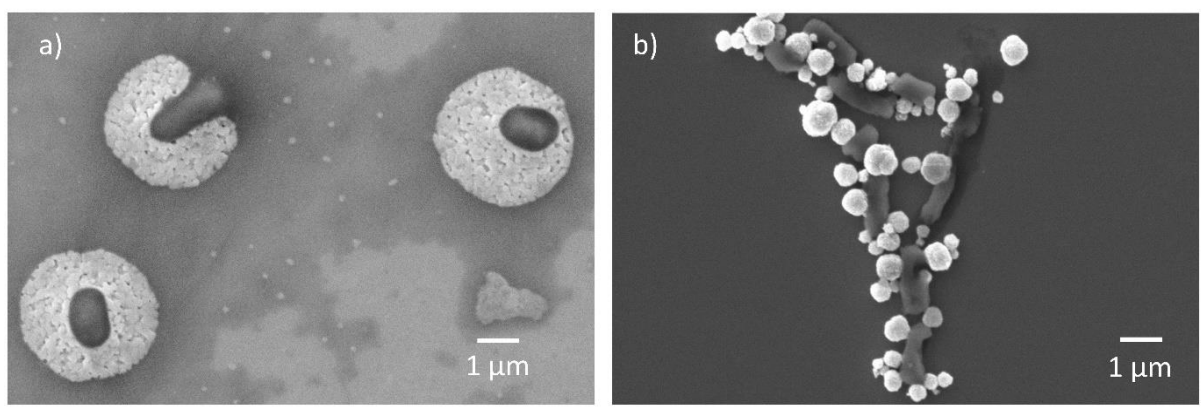

**Figure 8**. *SEM images of E. coli after incubation with (A) ZnO-OA-8 and (B) ZnO-HMTA-16.*

## 4. Discussion

The application of capping agents in the MW-assisted synthesis of NPs is a viable approach for tailoring and manipulating the physicochemical properties of the resulting NPs. Compared to conventional heating, the MW-assisted synthesis enables faster reaction rates, higher yields, and less energy, which reduces the costs and environmental impact [34] among other advantages [15]. We explored the use of two capping agents with different water interaction properties and the implications for their effectiveness in inhibiting bacterial growth. The primary aim of our comparative study was to synthesize ZnO NPs with opposite water wetting properties and study the implications on the physicochemical and antibacterial effects. Hydrophobic OA was chosen due to being previously used in MW-assisted synthesis of ZnO NPs for PLED device applications [14]. HMTA was selected as the hydrophilic counterpart because of its relative low cost and excellent solubility in polar solvents, which enables efficient growth of metal oxide nanostructures [35].

It has been reported in the literature that ZnO NPs differ in their efficacy at inhibiting bacteria growth depending on the strain of bacteria used. In some comparative studies that exposed different bacteria to the same ZnO NPs, Gram-positive bacteria were more sensitive towards exposure to ZnO NPs [36], [37]; yet in others they were less sensitive [38], [39]. Gram-negative bacteria (e.g., *E. coli*) typically exist as single cells that have a three-layered outer membrane structure [40], whereas Gram-positive bacteria (e.g., *S. aureus*) have a two-layered outer membrane structure [41]. Although Gram-positive bacteria have one less layer, they are structurally more rigid than Gram-negative bacteria because of a thicker peptidoglycan component that protects against physical damage. However, from **Figure 7**, it is clear that the most potent test sample (ZnO-HMTA-8) completely inhibited both Gram-negative *E. coli* and Gram-positive *S. aureus* at the same concentration (16 µg/mL), which suggests that the antibacterial mechanism was not affected by the bacteria's outer membrane structure. There was no trend observed for the remaining test samples with respect to the outer membrane structure. The lowest MIC value recorded from all test samples was 16 µg/mL from ZnO-HMTA-8 against both *E. coli* and *S. aureus*, which is comparable to the MIC value achieved using commercially available spherical 50 nm ZnO NPs against *E. coli* [38]. The same study also compared 10 µm ZnO particles and 'hedgehog' particles (needle clusters), and both ZnO types were less effective at inhibiting *E. coli* growth than the nanosized counterparts (MIC = 32 µg/mL), yet more effective than the remaining test samples from the current study. The MIC that was required to inhibit *S. aureus* growth was similar between the studies (MIC = 64 µg/mL), except for the uncapped ZnO and ZnO-OA-8 that required much greater concentrations to inhibit growth. In another study, ZnO nanorods doped with gallium achieved complete inhibition of *E. coli* growth at 128 µg/mL and *S. aureus* growth at 64 µg/mL, which was reduced to 64 µg/mL and 32 µg/mL after oxygen plasma treatment, respectively [37]. The improvement in the MIC after oxygen plasma treatment was associated with increased surface roughness caused by ion etching that resulted in a greater capability to damage the bacteria surface and prevent growth. In the following sections we thus detail other factors that are more prevalent in the antibacterial effect of ZnO NPs.

Since the presence of organic residues can hamper the activity of the NPs in practical applications [30], the EDX analysis was performed to establish whether the capping agents remained attached to the surface of the ZnO NPs after synthesis. Although residual carbon was detected on the surface of ZnO-OA (**Figure 3**), neither carbon nor nitrogen were found on the ZnO-HMTA (**Figure 3**). The fact that no organic residues were detected adhering to ZnO-HMTA suggests that MW-based synthesis is an effective technique to produce ZnO NPs that do not require post-synthesis purification. This could also explain why the typical absorbance peak of ZnO was more pronounced from ZnO-HMTA since the 'bare' nanoparticle surface would be able to interact with incident light more effectively than a surface covered with carbon residues (**Figure 5**). Despite this finding, the presence of HMTA-related residues should not be ruled out since the surface charge of both ZnO-HMTA was less positive than for the uncapped ZnO (**Figure 4**). This suggests that the surface of ZnO-HMTA was different from that of uncapped ZnO, but the concentration of C and N from HMTA remained undetected amongst the dominant signals from Zn and O in the EDX analysis. In the case of ZnO-OA (**Figure 4**), the surface charge measured from both test samples as colloids was negative, which could be a result of OA chemisorption onto the surfaces of ZnO NP producing negatively-charged carboxylate functional groups that has been previously reported [42]. In our experimental setup, we do not expect the capping agents themselves to have a major influence on the observed antibacterial effect. In the case of HMTA, it requires an acidic environment to undergo decomposition into more toxic byproducts (i.e., ammonia and formaldehyde) which then inactivate bacteria [43]. Since we performed our experiments in the Mueller Hinton Broth with a pH of 7.4, the acidic decomposition of the HMTA cannot contribute to the antibacterial activity of the tested ZnO NPs. OA is known to be inactive against Gram-negative species [44], yet it may exhibit antibacterial activity against Gram-positive bacterial species at MIC of 1 mg/mL. In our case, however, there is orders of magnitude lower concentration of the ZnO(OA) NPs. OA contribution to the antibacterial activity can thus be also neglected.

Both the primary particle size and surface charge of NPs are important properties when considering the mechanism(s) underlying the inhibition of bacteria growth. Ultra-small NPs that are typically in the single-digit nanometre size range (i.e., 1–9 nm) can traverse the bacterial membrane and become internalized, creating a toxic, unfavourable intracellular environment. The process whereby ultra-small NPs become internalized may occur without irreversible damage to the outer cell structure [45], but often some degree of damage is observed such as delamination of the outer membrane leading to cytoplasm leakage and loss of viability [46]. In specific cases [46], [47], it has been shown that ultra-small NPs can induce an antibacterial effect not through internalization but instead by interacting with the bacteria cell surface. This highlights the fact that NP size is not the sole property governing internalization and it emphasizes the complexity of elucidating the precise antibacterial mechanism of novel nanomaterials, including their comprehensive characterization in both powder and colloidal forms. However, particle size estimation from colloidal suspensions of test samples detected agglomerates hundreds of nanometres in diameter and did not detect any primary particles (**Figure 4**), which is in good agreement with high magnification images (**Figure 2A**). This suggests that the primary particle size of the test samples is not a predominant reason for the observed antibacterial effect. Also, the zeta potential value of the colloid appeared not to have a major role in the antibacterial effect. From zeta potential analysis (**Figure 4** A), the ZnO NPs with the largest value (above +30 mV) were the uncapped ZnO; yet this did not result in a larger antibacterial effect (**Figure 7**). The majority of bacteria possess an inherent net negative surface charge across the cell membrane [48], thus positively charged NPs have been shown to have a greater antibacterial effect than their negatively charged analogues due to electrostatic interactions [49]. The greatest antibacterial effect we observed was from the positively charged ZnO-HMTA with the lowest ZAD precursor concentration, i.e., ZnO-HMTA-8. Interestingly, uncapped ZnO also had a positive surface charge but recorded a higher MIC was less antibacterial than ZnO-HMTA. Negatively charged ZnO-OA was the least effective at inhibiting bacteria growth. While electrostatic interactions may account for the observed differences in the antibacterial activity between capping agents, the fact that uncapped ZnO was less efficient at inhibiting bacteria growth suggests that other mechanisms are involved. One such alternative pathway for ZnO NPs to inhibit bacterial growth is through the release of zinc ions ($Zn^{2+}$). $Zn^{2+}$ is essential in intracellular bacterial physiology and exists within a narrow concentration range maintained by homeostasis [50]; however, this can be disrupted by high extracellular concentrations from ZnO nanoparticle dissolution, which can lead to inward diffusion. High intracellular $Zn^{2+}$ concentrations can be cytotoxic and detrimental for bacterial growth [51], therefore $Zn^{2+}$ release from ZnO NPs could act as the extracellular $Zn^{2+}$ reservoir. However, **Figure 6** showed that the concentrations of $Zn^{2+}$ measured from the ZnO NP supernatants were too low to be cytotoxic towards bacteria, and, under these conditions, $Zn^{2+}$ release is not thought to be involved in the antibacterial mechanism of ZnO NPs. The lowest $Zn^{2+}$ released from capped ZnO was measured for ZnO-HMTA-8; therefore, $Zn^{2+}$ release also cannot be associated with the water affinity properties of the capping agent. However, since ZnO NPs had to be removed from solution for $Zn^{2+}$ measurement using the commercial (absorbance-based) assay, the absolute $Zn^{2+}$ concentration that interacts with bacteria could be understated [52]. It has been reported that $Zn^{2+}$ release from ZnO NP not removed from solution resulted in *E. coli* membrane damage that rendered the bacteria non-viable [53].

A major property of NPs that is fundamental to the interaction with other surfaces, such as the bacteria membrane, is the specific surface area (SSA), which was estimated using the BET method. **Figure 5** revealed that both ZnO-HMTA-capped variants adsorbed a greater amount of gas onto the surface than their OA-capped counterparts. The primary particle sizes of ZnO-OA-4 were similar to those of ZnO-HMTA-8, and ZnO-OA-8 similar to ZnO-HMTA-16 (**Figure 2A**), and all test samples experienced varying degrees of agglomeration as observed by SEM (**Figure 3A**, **3B**) and DLS analysis (**Figure 4B**). Therefore, the improvement in the antibacterial effect of HMTA-capped ZnO is likely a result of greater SSA and is not related to primary particle size or agglomeration characteristics, which is in good agreement with other reports that found an association between specific surface area and antibacterial activity of nanomaterials [54], including ZnO [31], [36].

Attempts to image the interaction between bacteria and test samples directly from broth immediately after the MIC test using SEM were unsuccessful due to the dense liquid, i.e., high OD values (**Figure 7**). Crystallization under vacuum of the salt content of the broth, alongside the bi-products of growth, prevented visualization of any bacteria or test samples. Therefore, we performed a parallel study to investigate the interaction between bacteria and test ZnO samples in a less complex liquid environment. The analysis revealed that ZnO interacted very differently with bacteria depending on the capping agent which is in good agreement with reported antibacterial effects of other capped ZnO [55]. Complete encapsulation of bacteria was observed by the hydrophobic ZnO-OA-8, which formed densely-packed islands, whereas bacteria were decorated with smaller agglomerates of hydrophilic ZnO-HMTA-16 (**Figure 8**). This different interaction behaviour can be in part explained by the hydrophobic nature of the *E. coli* bacteria surface [56], [57], [58], and superior interaction between similarly hydrophobic co-suspended particles such as ZnO-OA-8.

The positive surface charge of ZnO-HMTA in combination with the greater SSA could increase the probability of NP-bacteria interaction through electrostatic attraction coupled with a greater active surface area for interaction. Localised $Zn^{2+}$ release and possible reactive oxygen species (ROS) generation even without illumination [59], in addition to potential physical damage to the bacteria membrane upon NP interaction, all in combination would overwhelm bacterial defences and likely result in bacteria death. However, it is worth noting that, while both ZnO-HMTA samples inhibited *S. aureus* growth more effectively than their OA-capped counterparts (**Figure 7D**), ZnO-OA-4 inhibited *E. coli* growth more efficiently than ZnO-HMTA-16 (**Figure 7C**). Here, it is possible that the smaller size of nanoparticle agglomerates of ZnO-OA-4 as measured by DLS (**Figure 4B**) played a role in the efficient inhibition of *E. coli* growth. However, DLS analysis revealed the uncapped ZnO had the smallest particles of all test samples, which did not correlate with an enhancement of the antibacterial effect. Clearly, there are different modes of inactivation in the test samples, even though all were synthesized using the same technique. There is no single pathway responsible for the antibacterial effect of ZnO synthesized by the MW-based method in the presence/absence of capping agents, and it is likely dependent on the precise physicochemical properties of the synthesized NPs, which can vary greatly between different conditions of the synthesis environment.

## 5. Conclusions

Highly crystalline 10-20 nm ZnO NPs synthesized by MW-based method in the presence of different capping agents was an effective way to control surface charge and water interaction properties of the synthesized NPs. The smallest NPs with a positive surface charge synthesized with hydrophilic HMTA were the most effective at inhibiting the growth of both gram-positive *S. aureus* and gram-negative *E. coli* bacteria. However, ZnO NPs synthesized with hydrophobic OA with a negative surface charge still inhibited bacteria growth, but to a lesser extent despite their ability to encapsulate the bacteria cells. The results highlight that simple electrostatic interaction between the negatively charged bacteria and ZnO NP cannot be the main mechanism for the observed antibacterial effect, but more likely a combination of multiple factors working in tandem. We reveal that the specific surface area of ZnO NP is important when considering the underlying antibacterial mechanism, enabling greater interaction with the bacteria cell surface to impart localized $Zn^{2+}$ release, ROS generation, and physical damage to the bacteria membrane that culminates in the inhibition of bacteria growth.

## Acknowledgements

DR, BR and MŠB were funded by TAČR (TM03000033). In addition, BR and MŠB were funded by the European Union and Ministry of Education, Youth and Sports of the Czech Republic (CZ.02.01.01/00/22_008/0004596 - SenDiSo) and the Grant agency of the Czech Republic (24-10607J). MŠB would also like to acknowledge the student grant from CTU (SGS23/166/OHK4/3T/13). M.C. is grateful to the J. W. Fulbright Commission in the Czech Republic for the financial support through a postdoctoral fellowship (grant number: 2022-21-1) and further acknowledges the project DKRVO (RP/CPS/2024-28/007) supported by the Ministry of Education, Youth and Sports of the Czech Republic.

## CRediT authorship contribution summary

**DR:** conceptualization, methodology, investigation, data curation & formal analysis (zinc ion, UV-vis spectroscopy, MIC), writing (original & final version)
**MŠB:** investigation, data curation & formal analysis (zeta potential, DLS, MIC), writing (review & editing)
**TJ:** methodology & investigation (ZnO synthesis, XRD)
**PŠ:** investigation, data curation & formal analysis (SSA)
**MC:** conceptualization, methodology, investigation, data curation & formal analysis (TEM, SEM/EDX, SSA), funding acquisition, writing (original, review & editing)
**BR:** data analysis, funding acquisition, writing (review & editing)

## Data availability